\documentclass{article}
\usepackage{ijcai24}

\usepackage{times}
\usepackage{soul}
\usepackage{url}
\usepackage[hidelinks]{hyperref}
\usepackage[utf8]{inputenc}
\usepackage[small]{caption}
\usepackage{graphicx}
\usepackage{amsmath}
\usepackage{amsthm}
\usepackage{booktabs}
\usepackage{algorithm}
\usepackage{algorithmic}
\usepackage[switch]{lineno}
\usepackage{multirow}
\usepackage{float}
\usepackage{amsfonts}
\usepackage{mydef}

 \newcommand{\argmin}{\mathop{\arg\min}\hspace{0.3em}}

\title{From Principles to Applications: A Comprehensive Survey of Discrete Tokenizers in Generation, Comprehension, Recommendation, and Information Retrieval}

\author{
Jian Jia$^{1*}$
\and
Jingtong Gao$^{2*}$\and
Ben Xue$^1$\thanks{These authors contributed equally to this work.} \and
Junhao Wang$^1$\and
Qingpeng Cai$^1$\and
Quan Chen$^1$\and \\
Xiangyu Zhao$^2$\thanks{Corresponding author.}\and
Peng Jiang$^1$\And
Kun Gai$^1$ \\
\affiliations
$^1$Kuaishou Technology\\
$^2$City University of Hong Kong\\
\emails
\{jiajian, xueben, wangjunhao05, caiqingpeng, chenquan06, jiangpeng, yuyue06\}@kuaishou.com,
jt.g@my.cityu.edu.hk,
xianzhao@cityu.edu.hk
}

\begin{document}

\maketitle

\begin{abstract}

Discrete tokenizers have emerged as indispensable components in modern machine learning systems, particularly within the context of autoregressive modeling and large language models (LLMs). 
These tokenizers serve as the critical interface that transforms raw, unstructured data from diverse modalities into discrete tokens, enabling LLMs to operate effectively across a wide range of tasks. 
Despite their central role in generation, comprehension, and recommendation systems, a comprehensive survey dedicated to discrete tokenizers remains conspicuously absent in the literature. 
This paper addresses this gap by providing a systematic review of the design principles, applications, and challenges of discrete tokenizers. 
We begin by dissecting the sub-modules of tokenizers and systematically demonstrate their internal mechanisms to provide a comprehensive understanding of their functionality and design.
Building on this foundation, we synthesize state-of-the-art methods, categorizing them into multimodal generation and comprehension tasks, and semantic tokens for personalized recommendations. 
Furthermore, we critically analyze the limitations of existing tokenizers and outline promising directions for future research. 
By presenting a unified framework for understanding discrete tokenizers, this survey aims to guide researchers and practitioners in addressing open challenges and advancing the field, ultimately contributing to the development of more robust and versatile AI systems.

\end{abstract}

\section{Introduction}


\begin{figure*}
    \centering
    \includegraphics[width=0.95\linewidth]{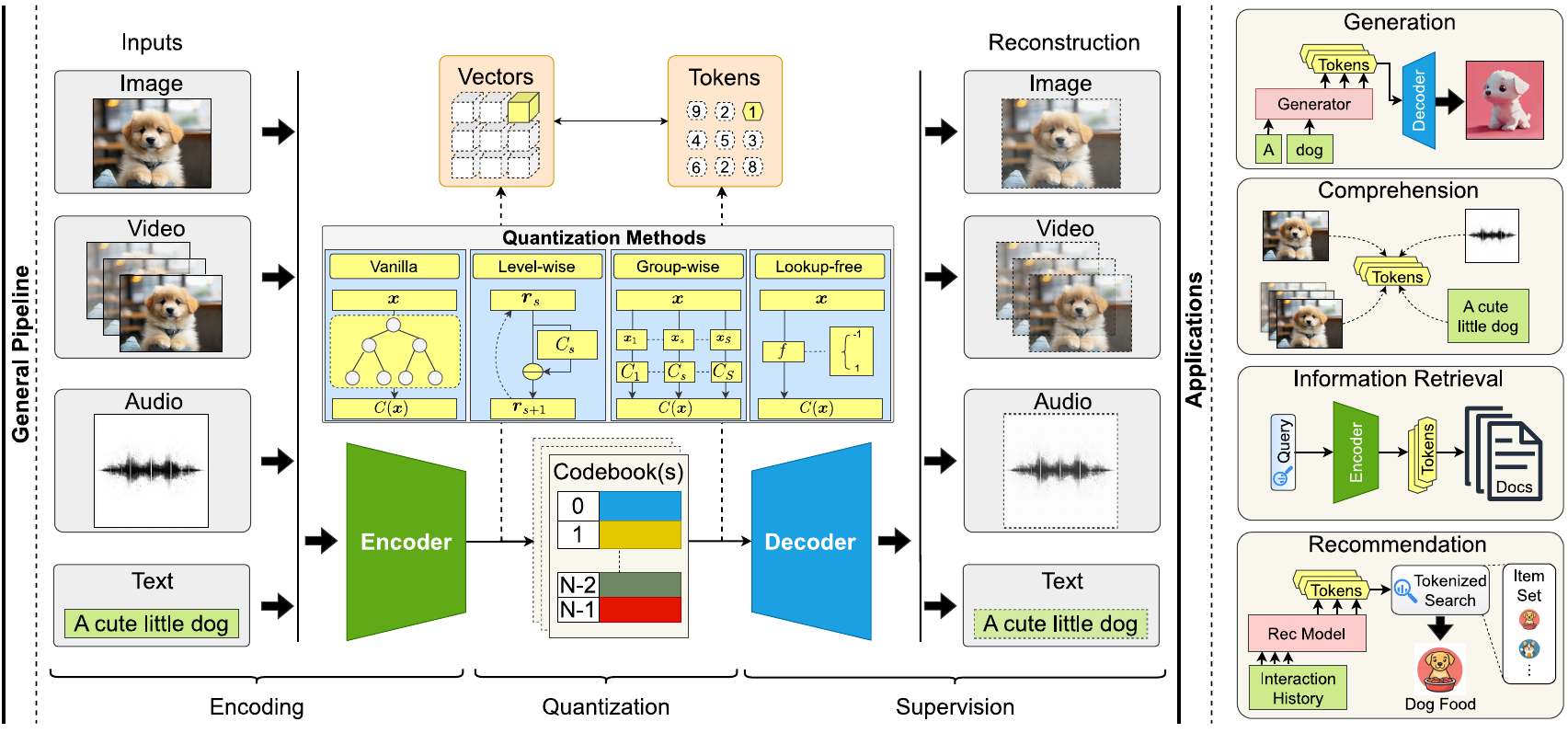}
    \caption{Illustration of general pipeline and applications. Discrete tokenizers can be regarded as the \textbf{interface} between different input modalities and downstream tasks.}
    \label{fig:enter-label}
    \vspace{-3mm}
\end{figure*}

The rapid advancement of AI systems and Large Language Models (LLMs) has underscored the critical role of tokenizers as indispensable components in modern machine learning systems. Tokenizers serve as the bridge between raw, unstructured data from diverse modalities 
and the discrete tokens that LLMs process. By transforming continuous or heterogeneous inputs into a sequence of discrete symbols, tokenizers enable LLMs to operate effectively across a wide range of tasks. As AI continues to evolve, the ability of LLMs to generate and comprehend complex information hinges significantly on the quality and design of their tokenization mechanisms. In this survey, we focus specifically on discrete tokenizers, which are uniquely suited for integration with LLMs due to their ability to discretize information while preserving semantic fidelity \cite{team2024chameleon,chen2024next}.

Tokenizers play multifaceted roles in various domains, including generation, understanding, and recommendation tasks. 
In generative applications, tokenizers decompose input signals (e.g., pixels in videos or words in text) into discrete tokens. Tokenizers enable autoregressive modeling for video synthesis \cite{tian2024visual} and text generation \cite{guo2025deepseek}, determining how well an LLM can produce coherent and contextually relevant outputs.
For comprehension tasks, tokenizers~\cite{team2024chameleon,chen2025janus} map raw modality inputs into the unified token format, directly influencing the token-based multimodal models' capacity to understand different modalities.
In recommender systems, tokenizers~\cite{rajput2023tiger,singh2024better} transform semantic embedding of user preferences and item attributes into semantic IDs, improving the generalization of personalized recommendations in short-video platforms and online shopping.
Despite their pivotal role and pervasive importance, there remains a notable gap in the literature: no comprehensive survey has systematically reviewed the design, application, and challenges of discrete tokenizers across these tasks.
A thorough examination of discrete tokenizers is essential not only to trace the evolutionary trajectory and provide an overview of SOTA methods but also to highlight their current limitations and guide future research efforts toward addressing these challenges.

The significance of surveying discrete tokenizers cannot be overstated. 
First, generation and comprehension tasks form the cornerstone of AI applications, while recommender systems are integral to modern digital economies, powering applications such as e-commerce and content delivery platforms. 
Second, state-of-the-art approaches in generation \cite{chen2024next}, comprehension~\cite{yang2023teal}, and recommendation tasks \cite{li2023large} rely heavily on the capabilities of LLMs, which necessitate the discretization of input data through tokenizers. 
Third, tokenizers act as a universal interface that harmonizes different modalities (text, image, video, audio) \cite{guo2025deepseek,tian2024visual,tan2024sweettokenizer,zhang2023speechtokenizer} at both syntactic and semantic levels, ensuring seamless interaction between raw data and LLMs. 
Finally, the performance ceiling of any LLM-based system is inherently constrained by the quality of its tokenizer; suboptimal tokenization can lead to significant degradation in downstream task outcomes \cite{yang2023teal,yu2023magvit-v2}. Given these considerations, a detailed exploration of discrete tokenizers is imperative to advance our understanding of their impact and potential.

Our survey differs significantly from existing reviews in the field and aims to provide a comprehensive framework for understanding semantic tokenizers in contemporary AI systems. While previous works have focused on specific aspects -- traditional vector quantization techniques for compression~\cite{wu2019vector}, quantization methods in recommender systems~\cite{liu2024vector}, and VQ-VAE variants for representation learning~\cite{zheng2023online} -- this survey examines semantic tokenizers through the perspective of modern neural architectures and their role in enabling unified multimodal processing applications. 
Given the increasing centrality of tokenization~\cite{wang2024tokenization} in foundation LLMs and the rapid emergence of novel multimodal architectures, a thorough analysis of semantic tokenization techniques becomes essential for guiding future research directions. 

This survey is structured to provide a comprehensive overview of discrete tokenizers from the perspectives of mechanisms, applications, and challenges.
In Sec.~\ref{sec:mechanism}, we begin by formally defining tokenizers and delineating the key components involved in their design, with a particular emphasis on discrete quantization.
Sec.~\ref{sec:application} explores the practical applications of tokenizers in generation, comprehension, recommendation, and information retrieval tasks, highlighting the unique contributions of SOTA methods in each domain. 
In Sec.~\ref{sec:challenge}, we critically examine the current challenges and open problems associated with discrete tokenizers, offering insights into promising directions for future research. 
Through this systematic review, we aim to equip researchers and practitioners with a deeper understanding of discrete tokenizers and inspire innovations that address existing limitations in the field.

\section{Taxonomy of Discrete Tokenizer} \label{sec:mechanism}

\begin{table*}[h!]
\centering
\resizebox{0.95\textwidth}{!}{
\begin{tabular}{l|c|c|c|c}
\toprule
\textbf{Methods} & \textbf{Modality} & \textbf{Backbone} & \textbf{Quantizer} & \textbf{Task} \\ \midrule
VQVAE~\cite{van2017vqvae} & Image & CNN & VQ & Generation \\
VQGAN~\cite{esser2021vqgan} & Image & CNN & VQ & Generation \\
ViT-VQGAN~\cite{yu2021vit-vqgan} & Image & Transformer & VQ & Generation \\
RQVAE~\cite{lee2022rqvae} & Image & CNN & RQ & Generation \\
LQAE~\cite{liu2024lqae} & Image & CNN & VQ & Generation \\
SEED~\cite{ge2023seed} & Image & Transformer & VQ & Generation, Comprehension \\
TiTok~\cite{yu2024titok} & Image & Transformer & VQ & Generation \\
MAGVIT~\cite{yu2023magvit} & Image, Video & CNN & VQ & Generation \\
MAGVIT-v2~\cite{yu2023magvit-v2} & Image, Video & CNN & LFQ & Generation \\
OmniTokenizer~\cite{wang2024omnitokenizer} & Image, Video & Transformer & VQ & Generation \\
SweetTokenizer~\cite{tan2024sweettokenizer} & Image, Video & Transformer & VQ & Generation, Comprehension \\
Cosmos~\cite{agarwal2025cosmos} & Video & CNN, Transformer & FSQ & Generation \\
VidTok~\cite{tang2024vidtok} & Video & CNN & FSQ & Generation \\
TEAL~\cite{yang2023teal} & Image, Audio, Text & Transformer & VQ & Comprehension \\
AnyGPT~\cite{zhan2024anygpt} & Image, Audio, Text & Transformer & VQ & Generation, Comprehension \\
LaViT~\cite{jin2024lavit} & Image & Transformer & VQ & Generation \& Comprehension \\
Video-LaViT~\cite{jin2024video-lavit} & Video & Transformer & VQ & Generation \& Comprehension \\
ElasticTok~\cite{yan2024elastictok} & Image, Video & Transformer & VQ, FSQ & Generation, Comprehension \\
Chameleon~\cite{team2024chameleon} & Image, Text & CNN, Transformer & VQ & Generation, Comprehension \\
ShowO~\cite{xie2024show} & Image, Text & CNN, Transformer & LFQ & Generation, Comprehension \\
SoundStream~\cite{zeghidour2021soundstream} & Audio & CNN & RQ & Generation \\
iRVQGAN~\cite{kumar2024improvedRVQGAN} & Audio & CNN & RQ & Generation \\
HiFiCodec~\cite{yang2023hificodec} & Audio & CNN & GRVQ & Generation \\
RepCodec~\cite{huang2023repcodec} & Audio & CNN, Transformer & RQ & Comprehension \\
SpeechTokenizer~\cite{zhang2023speechtokenizer} & Audio & CNN, Transformer & RQ & Generation, Comprehension \\
NeuralSpeech-3~\cite{ju2024naturalspeech} & Audio & CNN, Transformer & VQ & Generation, Comprehension \\
TIGER~\cite{rajput2023tiger} & Text & MLP & RQ & Recommendation \\
SPM-SID~\cite{singh2024better} & Text & MLP & RQ & Recommendation \\
TokenRec~\cite{qu2024tokenrec} & Text & MLP & VQ & Recommendation \\
VQ-Rec~\cite{hou2023learning} & Text & MLP & PQ & Recommendation \\
LC-Rec~\cite{zheng2024adapting} & Text & MLP & RQ & Recommendation \\
LETTER~\cite{wang2024learnable} & Text & MLP & RQ & Recommendation \\
CoST~\cite{zhu2024cost} & Text & MLP & RQ & Recommendation \\ 
ColaRec~\cite{wang2024content} & Text & MLP & VQ & Recommendation \\
SEATER~\cite{si2024generative} & Text & MLP & VQ & Recommendation \\
QARM~\cite{luo2024qarm} & Text & MLP & VQ & Recommendation  \\
DSI~\cite{tay2022transformer} & Text & Transformer & VQ & Information Retrieval \\
Ultron~\cite{zhou2022ultron} & Text & Transformer & PQ & Information Retrieval \\
GenRet~\cite{sun2024learning} & Text & Transformer & VQ & Information Retrieval \\
LMINDEXER~\cite{jin2023language} & Text & Transformer & VQ & Information Retrieval \\
RIPOR~\cite{zeng2024scalable} & Text & Transformer & RQ & Information Retrieval \\
\bottomrule
\end{tabular}
}
\vspace{-3mm}
\caption{
Taxonomy and summarization of discrete tokenizers across various modalities for tasks such as generation, comprehension, recommendation and information retrieval. 
It highlights the use of different quantization strategies and model architectures in addressing diverse tasks.
}
\vspace{-3mm}
\end{table*}

\subsection{General Pipeline}
\label{sec:procedure}

The concept of discrete tokenizer originates from natural language processing tasks, which aims to break down the input into several finest tokens from a pre-defined dictionary~\cite{bpe2016}.
However, for modalities containing much denser information (e.g.\ image, video, audio), directly tokenizing raw input is not feasible. 
Therefore, a standardized process must be established to discretize various modalities universally.
To this end, we summarize the discrete tokenization pipeline, which consists of three steps:

\paragraph{(1) Encoding.}
An encoder $Enc$ is required to map input data $\boldsymbol{x}$ from a lower-dimensional tensor to a higher-dimensional latent vector $\boldsymbol{z} \in \mathbb{R}^{d}$:
\begin{equation}\label{equ:encoding}
\begin{gathered}
    \boldsymbol{z} = Enc(\boldsymbol{x})
\end{gathered}
\end{equation}

This transformation allows the model to capture complex patterns and relationships within the data. 
By leveraging various architectures, such as neural networks or transformers, the encoder learns to represent the input in a space that highlights relevant features.

\paragraph{(2) Quantization.}
Given a set of continuous vectors $\boldsymbol{z} \in \mathbb{R}^d$ forming a matrix $\boldsymbol{Z} \in \mathbb{R}^{n\times d}$, vector quantization aims to learn a codebook matrix $\boldsymbol{C} \in \mathbb{R}^{m\times d}$ containing $m$ discrete codes, where typically $m \ll n$. The quantization process can be formalized as a mapping function $Q: \boldsymbol{Z} \rightarrow \boldsymbol{C}$ that assigns each input vector $\boldsymbol{z}$ to its nearest code $\boldsymbol{c}_j$ in the codebook:

\begin{equation}\label{equ:quantization}
\begin{gathered}
    [j, \boldsymbol{c}_j] = Q(\boldsymbol{z})\\
    j = \underset{k=1,...,m}{\argmin} D(\boldsymbol{z}, \boldsymbol{c}_k),
\end{gathered}
\end{equation}
where $D$ measures the distance between vectors, $j$ is the quantized codeword, and $\boldsymbol{c}_j$ is the quantized representation.

\paragraph{(3) Supervision.}
A decoder $Dec$ is adopted to reconstruct the original input from the discretized representation $\boldsymbol{c}$. 
\begin{equation}\label{equ:decoding}
\begin{gathered}
    \hat{\boldsymbol{x}} = Dec(\boldsymbol{c})
\end{gathered}
\end{equation}

By minimizing the reconstruction error between the original input $\boldsymbol{x}$ and its reconstructed version $\hat{\boldsymbol{x}}$, the model is encouraged to preserve important information in the tokenized form.
\begin{equation}
\mathcal{L}_{rec}=\left\|\boldsymbol{x}-\boldsymbol{\hat{x}}\right\|_2^2
\end{equation}

Straight-Through Estimator (STE)~\cite{van2017vqvae} allows gradient flow through the discrete quantization operation and the codebook to be learned and updated automatically. 
STE simply copies the gradients directly:

\begin{equation}
\frac{\partial \boldsymbol{c}}{\partial \boldsymbol{z}} \approx \frac{\partial \boldsymbol{z}}{\partial \boldsymbol{z}}=\mathbf{I}
\end{equation}

Incorporated with commitment loss $\mathcal{L}_{cmt}$, the whole process is differentiable and the encoder output is limited around codebook vectors:
\begin{equation}
\mathcal{L}_{cmt}=\left\|\operatorname{sg}\left[\boldsymbol{z}\right]-\boldsymbol{c}\right\|_2^2+\left\|\boldsymbol{z}-\operatorname{sg}[\boldsymbol{c}]\right\|_2^2,
\end{equation}
where $\operatorname{sg}$ stands for the stopgradient operator.

Through the tokenization process, the inputs with similar contents are mapped to closer discrete, low-dimensional tokens. 
As a result, these quantized representations, characterized by their semantic, discrete, and low-dimensional properties, are increasingly employed in token-prediction models, particularly large language models (LLMs).

\subsection{Backbone Network}

\paragraph{MLP-based.}
MLP-based approaches were commonly used in early stages. However, in recent years, many methods in the recommendation domain have also adopted MLP as the backbone. 
This shift is primarily due to the fact that in recommender systems, each entity (e.g.\ user, item) is typically represented by an embedding vector. 
To effectively model and process these embeddings, MLP is employed to perform latent space mapping~\cite{rajput2023tiger,singh2024better}, enabling more sophisticated feature interactions and enhancing recommendation performance.

\paragraph{CNN-based.}
In the context of image modalities, CNN-based network architectures, such as UNet~\cite{ronneberger2015unet}, are widely adopted as the backbone for feature extraction~\cite{esser2021vqgan}. 
Similarly, for audio modalities, CNNs can be effectively utilized by converting audio signals into spectrograms~\cite{zeghidour2021soundstream}, allowing the network to process the spectral features. 

When handling even higher-dimensional modalities, such as video, these architectures are extended to 3D-CNNs
, which incorporate an additional temporal dimension alongside the spatial dimensions, enabling the capture of both spatial and temporal information for more comprehensive analysis~\cite{yu2023magvit}.

\paragraph{Transformer-based.}
The development of attention mechanisms and transformer models~\cite{dosovitskiy2020vit} has demonstrated significant enhance performance across various tasks. 
For text information, we can use the transformer architecture to extract relevant information and perform sequence-to-sequence reconstruction~\cite{sun2024learning}.
For 2D inputs, such as images, the data is typically divided into patches, and these patches are then treated as a sequence, fed into the transformer's encoder-decoder framework, as seen in~\cite{yu2021vit-vqgan}. 
For 3D inputs, such as videos, the data is sliced along the temporal dimension, converting 2D patches into 3D tubes, as seen in~\cite{wang2024omnitokenizer}.
The key advantage of transformers lies in their scalability, allowing for the efficient handling of large parameter sizes. As a result, many of the latest state-of-the-art methods in various fields now adopt transformers as the backbone to leverage their ability to capture complex data features~\cite{yu2024titok}.

More recently, the transformer-based structure has also been adopted to reduce token number through cross-attention querying.
For example, SweetTokenizer~\cite{tan2024sweettokenizer} adopts the approach of Q-Former~\cite{li2023blip} by learning a set of fixed input queries to reduce redundant information embedded in neighbor patches.

\subsection{Quantization Method}

\paragraph{Vanilla Quantization.}
Pioneered by Shannon et al.~\cite{shannon1959coding}, vanilla vector quantization encompasses methods that construct codebooks with explicit structures such as clusters~\cite{nister2006scalable} or trees~\cite{babenko2015tree}. 
The quantization procedure can be described by Eq~\eqref{equ:quantization}.

Recent works argue that lexical tokens inherently contain higher-level semantic information, and thus, the latent space of text tokens can come to the rescue for image discretization.
For example, LQAE~\cite{liu2024lqae} quantizes image embeddings using a pretrained language codebook (e.g., BERT~\cite{devlin2018bert}), aligning images and text without paired data, thus enabling few-shot image classification with LLM.

\paragraph{Level-wise Quantization.}
Vanilla Quantization often introduces rough errors during the quantization process, raising the challenge of improving the approximation of vectors using a codebook. 

One possible solution is the level-wise quantization approach.
This method suggests that after quantizing at each level, the codebook from the subsequent level is used to approximate the quantization error of the current level, as seen in RQ~\cite{juang1982multiple}. 
It employs a new sequential quantization process where each stage k quantizes the residual from previous stages:

\begin{equation}\label{equ:rvq}
\boldsymbol{r}_{s+1} = \boldsymbol{r}_{s} - C_s(\boldsymbol{r}_s)
\end{equation}
where $\boldsymbol{r}_s$ denotes the $s_{th}$ stage of residual vector with $\boldsymbol{r}_0=\boldsymbol{x}$, $C_s \in C$ corresponds to the $s_{th}$ codebook, and a quantization result of $[C_0(\boldsymbol{r}_0), ..., C_s(\boldsymbol{r}_s), ..., C_S(\boldsymbol{r}_S)]$ could be obtained with a total of $S$ codewords instead of a single codeword. 
By iteratively accumulating these corrections across multiple levels, the overall approximation becomes more precise, leading to a reduction in quantization error and enhancing the quality of the results. This approach has been explored as an effective means to refine vector quantization.

RQ adopts a greedy approach by choosing the nearest neighbor within the current layer, but this method doesn't guarantee a global optimum. To overcome this, Additive Quantization (AQ)~\cite{babenko2014additive} improves upon it by using beam search to aggregate one selected code per codebook, representing vectors as sums of $S$ codewords, and applying joint optimization for enhanced performance:

\begin{equation}
\boldsymbol{x} \approx \sum_{s=1}^S C_s(\boldsymbol{x})
\end{equation}

\paragraph{Group-wise Quantization.}
Another approach, known as group-wise quantization, suggests that splitting a vector into multiple subcomponents and quantizing each part separately can also help reduce the quantization error. By handling smaller portions of the vector independently, this method allows for more accurate representation and finer control over the quantization process. 
For example, Product Quantization (PQ)~\cite{jegou2010product} established another parallel approach by decomposing high-dimensional vectors into $S$ separately orthogonal quantized subvectors:

\begin{equation}
C(\boldsymbol{x}) = (C_1(\boldsymbol{x}_1), ..., C_s(\boldsymbol{x}_s), ..., C_S(\boldsymbol{x}_S)),
\end{equation}
where $\boldsymbol{x} = \text{concat}(\boldsymbol{x}_1, ..., \boldsymbol{x}_s, ..., \boldsymbol{x}_S)$. This decomposition also enables efficient distance computation through smaller lookup tables. 

Optimized Product Quantization (OPQ)~\cite{ge2013optimized} extends this by learning an optimal rotation matrix R that minimizes quantization distortion while maintaining subspace independence. 

\paragraph{Lookup-free Quantization.}
The methods discussed above typically involve maintaining a codebook and performing lookups. However, recent research~\cite{yu2023magvit-v2} has shown that as the codebook size increases, optimization becomes more challenging. This is due to the fact that many entries in a larger codebook remain unused, which can hinder the optimization process and degrade performance.
As a result, some approaches have been proposed that eliminate the need for lookup tables or traditional codebooks. 

Finite Scalar Quantization (FSQ) \cite{mentzer2023fsq} projects the 
encoded representation to a few dimensions via transformation $f$, with each dimension rounded to a small set of fixed values, forming an implicit codebook:
\begin{equation}
    Q(\boldsymbol{z}) = \text{round}(f(\boldsymbol{z}))
\end{equation}

By carefully bounding each channel, FSQ can create a codebook of any desired size. 
For a vector \( \boldsymbol{z} \) with \( d \) channels, mapping each entry \( \boldsymbol{z}_i \) to \( L \) values (e.g., \( \boldsymbol{z}_i \mapsto \lfloor L/2 \rfloor \tanh(\boldsymbol{z}_i) \) followed by rounding) results in  \( L^d \) possible vectors.

More aggressively, Lookup-Free Quantization (LFQ)~\cite{yu2023magvit-v2} decomposes the latent space into binary dimensions, and each dimension is quantized independently with $\boldsymbol{c}_i = \{-1,1\}$. 
The latent space of LFQ is decomposed as the Cartesian product \( C = \prod_{i=1}^{\log_2 K} C_i \). Given \( \boldsymbol{z} \in \mathbb{R}^{\log_2 K} \),
The \( \arg \min \) can be computed using the sign function as:
\begin{equation}
Q(\boldsymbol{z}_i) = \text{sign}(\boldsymbol{z}_i) = -1 \cdot (\boldsymbol{z}_i \leq 0) + 1 \cdot (\boldsymbol{z}_i > 0).
\end{equation}

Experiments show that LFQ can grow the vocabulary size in a way
benefiting the generation quality of language models.

\section{Applications} \label{sec:application}

Discrete tokenization was initially widely adopted in the field of Natural Language Processing (NLP) for segmenting sentences into subwords \cite{bpe2016,wu2016google,kudo2018sentencepiece}. Building on this foundation, tokenization techniques have since been extended to non-textual modalities, garnering significant attention and achieving substantial advancements \cite{team2024chameleon,xie2024show,chen2025janus}. These developments have paved the way for constructing fully token-in-token-out multimodal large language models \cite{yin2023survey}, enabling seamless integration of diverse data types within a unified framework. 
Meanwhile, discrete tokenizers applied in LLM-based generative recommendation models \cite{li2023large} have gradually garnered attention as a means to complement or even replace traditional ID features in conventional recommender systems.

In this section, we discuss the recent advances of discrete tokenizers applied in generation, comprehension, recommendation, and information retrieval, and propose a unified framework to summarize and categorize the existing studies systematically.

\subsection{Generation}

Tokenizers, exemplified by Byte Pair Encoding (BPE), were initially widely adopted in the field of generation task of NLP to transform sentences into discrete tokens, thereby aligning with the requirements of transformer-based backbones in large language models (LLMs).

Inspired by discrete representations in NLP, VQ-VAE \cite{van2017vqvae} and VQ-GAN \cite{esser2021vqgan} employ vector quantization (VQ) to partition raw pixels into a fixed number of tokens as the pioneers in image generation. To reduce the sequence length of the quantized feature map and improve the image fidelity for autoregressive modeling, \cite{lee2022rqvae} introduces the residual quantization (RQ) mechanism \cite{juang1982multiple} to represent the latent space in a residual way. 
To overcome the limitations of 2D physical patch positions in prior tokenizers, SEED \cite{ge2023planting} introduces a VQ-based image tokenizer that generates discrete visual codes with 1D causal dependency and high-level semantic representations.
As a contemporaneous work, TiTok \cite{yu2024titok} introduces the 1-dimensional tokenizer to learn the compact latent tokens independent of the input resolution and reduce the inherent redundancy in adjacent image pixels. To facilitate the training efficiency, a two-stage training strategy including supervision with proxy code and decoder fine-tuning is also proposed.

MAGVIT proposes the 3D-VQ architecture to construct the spatial-temporal tokenizer, which models the temporal dynamics in the video generation task. To further improve the 
generation quality and solve the codebook collapse of the VQVAE, MAGVITv2 \cite{yu2023magvit-v2} propose the Lookup-Free Quantization (LFQ) as the replacement of the VQ. It also improves the tokenizer's encoder and decoder architecture, such as causal 3D-CNN and sampler operation, to achieve a visual tokenizer capable of tokenizing images and videos using a shared codebook.
Instead of using 3D-CNN as the backbone, Omnitokenizer \cite{wang2024omnitokenizer} adopts the spatial-temporal decoupled transformer and a progressive training schedule to joint image and video tokenization.
To further break the limitation of the rasterization order in 2D patches and achieve a higher compression ratio, SweetTokenizer \cite{tan2024sweettokenizer} encodes the original image/frame patches into the token sequence with the cross-attention query auto-encoder (CQAE), significantly reducing the number of tokens required for discretization.
Instead of using LFQ like MAGVITv2, Cosmos \cite{agarwal2025cosmos} and VIDTOK \cite{tang2024vidtok} introduce the Finite Scalar Quantization (FSQ) \cite{mentzer2023fsq} to generate discrete tokens, where each dimension is quantized to a small, fixed set of values.

Audio generation also faces the same challenge as image generation, which is how to compress the original signal into low bit-rate streams with high fidelity. 
SoundStream~\cite{zeghidour2021soundstream} adopts a fully causal convolution architecture in tokenizer's backbone structure, which ensures that the network encodes and decodes audio signals based solely on previous samples.
iRVQGAN~\cite{kumar2024improvedRVQGAN} decouples code lookup and code embedding by performing code lookup in a low-dimensional space, while the code embeddings reside in a higher-dimensional space. Additionally, it applies L2-normalization to the codebook vectors, converting Euclidean distance into cosine similarity, which improves both stability and quality.
HiFi-Codec \cite{yang2023hificodec} adopts grouped residual vector quantization (GRVQ) to address the issue in RQ models, where the first layer contains most of the information while subsequent layers contain less.

\subsection{Comprehension}

Autoregressive (AR) large language models have ushered in a new era in the field of artificial intelligence, marking a significant milestone in the advancement of generative and comprehension capabilities. Tokenizers, discretizing any modality into a token sequence, facilitate the development of universal interfaces that bridge diverse data types and enable seamless interaction with transformer-based LLM.
TEAL \cite{yang2023teal} and AngGPT \cite{zhan2024anygpt} adopt modality-specific tokenizers to discretize image, text, and audio information into one discrete token sequence for token-in-token-out multimodal LLM.
To solve the restriction of encoding images/videos to a fixed number of tokens irrespective of the original visual content, 
LaViT \cite{jin2024lavit} and ElasticTok \cite{yan2024elastictok} generate dynamic discrete visual tokens maintaining high-level semantics. 
With the tokenizer as the universal interfaces of text and visual modalities and large-scale pretraining, Chameleon \cite{team2024chameleon} and Show-O \cite{xie2024show} demonstrate the broad and general capabilities and achieve promising performance in various downstream applications of generation and comprehending tasks.

As for audio comprehension, recent works increasingly focus on reconstructing audio signals from multiple informational perspectives, not only emphasizing the fidelity of local detail reconstruction but also considering the semantic information of the discretized tokens.
RepCodec~\cite{huang2023repcodec} learns a vector quantization codebook by reconstructing speech representations from speech encoders like HuBERT~\cite{hsu2021hubert}.
Furthermore, SpeechTokenizer~\cite{zhang2023speechtokenizer} unifies both semantic and acoustic tokens, hierarchically disentangling different aspects of speech information across multiple RQ layers, which boosts the alignment of speech and language tokens.
More comprehensively, NeuralSpeech-3~\cite{ju2024naturalspeech} introduces a factorized neural speech codec that decomposes complex speech waveforms into separate subspaces representing attributes like content, prosody, timbre, and acoustic details, and then reconstructs high-quality speech from these factors.

\subsection{Recommendation}

Existing recommender systems (RS) predominantly rely on unique IDs randomly assigned to users and items, such as ItemIDs or UserIDs. The ID representation lacks inherent semantic information and heavily relies on extensive historical interactions to learn the ID embeddings.
Consequently, the acquired recommendation model struggles to generalize to unseen data, resulting in suboptimal performance, particularly in cold-start scenarios and long-tail user recommendations. Notably, in the recommendation domain, semantic IDs are commonly used to represent discrete semantic tokens for items or users. Therefore, we use semantic ID and semantic token interchangeably in this part.

To address these critical limitations of traditional ID-based approaches, integrating semantic IDs into recommender systems has garnered increasing attention from both industries and academics.
First, semantic IDs offer a robust solution to the challenges of data sparsity, cold-start problems, and long-tail distribution. Unlike conventional IDs, which rely solely on interaction data (e.g., user clicks), semantic IDs leverage content semantics -- such as textual descriptions or visual features --to generate meaningful initial representations for new users or items, thereby mitigating the above issue. 
Second, semantic IDs facilitate cross-domain and cross-modal recommendations by acting as a bridge between different domains, enabling knowledge transfer and improving generalization across heterogeneous data sources. 
Finally, semantic IDs align seamlessly with the input requirements of large language models (LLMs), enabling the seamless integration of these identifiers into LLM-driven systems. This compatibility not only facilitates the inheritance of key properties of LLMs, such as scaling laws, but also allows recommendation systems to benefit from the continuous improvements in model capacity and generalization that arise from scaling up data and model size. By leveraging the intrinsic connection between semantic IDs and LLMs, recommendation systems can achieve enhanced performance, scalability, and adaptability, paving the way for more robust and intelligent personalized recommendation paradigms.

As the pioneer, TIGER~\cite{rajput2023tiger} first proposes the generative recommendation framework, including the content embedding generator, semantic ID tokenizer, and Transformer-based autoregressive recommender. The content embedding is extracted from the item text content by the text encoder. The core component -- semantic ID tokenizer -- is based on RQVAE, which is introduced to quantize the content embedding into sequential semantic tokens with a hierarchical structure. Item semantic tokens are chronologically organized to the user historical interaction sequence as the input of the autoregressive recommender. Inspired by the success in the academic datasets, \cite{singh2024better} applies the semantic tokens generated by TIGER and replaces the original video ID in the real-world ranking model of the YouTube platform. The online industry results show that the semantic ID scheme improves the generalization in cold-start items and achieves comparable performance in the overall items.

For incorporating high-order collaborative knowledge in LLM-based recommendations, TokenRec \cite{qu2024tokenrec} proposes a masked vector-quantized tokenizer (MQ-Tokenizer), which quantizes collaborative embeddings of user and item into discrete tokens. The MQ-Tokenizer randomly masks the input collaborative embeddings to enhance the token generalization capability.
To enhance the transferability of sequence modeling and improve its performance in cross-domain scenarios, VQ-Rec \cite{hou2023learning} quantize the BERT embedding of item text based on product quantization (PQ) and conduct code-embedding alignment to achieve transfer capacity in downstream domains.
Instead of utilizing unstructured VQ and PQ, LC-Rec \cite{zheng2024adapting}, LETTER \cite{wang2024learnable}, and CoST \cite{zhu2024cost} adopts the RQ mechanism to create discrete hierarchical item tokens based on LLaMA embedding~\cite{touvron2023llama}.
Different from the above methods, ColaRec \cite{wang2024content}, SEATER \cite{si2024generative}, and QARM \cite{luo2024qarm} use k-means algorithm to conduct hierarchical clustering and generate tree-structured item tokens from the content or collaborative embedding.

\subsection{Information Retrieval}

With the advancement of pre-trained language models, generative information retrieval \cite{li2024matching} has emerged as a novel paradigm, gaining increasing attention in recent years. For the sake of terminological consistency, the concept of doc identifiers in the Information Retrieval (IR) domain is also referred to as the doc token in the following discussion.

DSI \cite{tay2022transformer} explores the semantically structured doc token by conducting the hierarchical k-means algorithm on doc embeddings, allowing semantically similar documents to share the same prefix tokens.
Ultron \cite{zhou2022ultron} introduces the PQ mechanism to compress the dense vector space from the pre-trained encoder into compact discrete token space.
GenRet~\cite{sun2024learning} and LMINDEXER~\cite{jin2023language} utilize the encoder-decoder transformer to autoregressively generate the semantic continuous representation, which is quantized into discrete tokens based on maximum inner-product with learnable codebook tables.
To further minimize the distortion error between the original and approximated representations, RIPOR \cite{zeng2024scalable} adopts the RQ mechanism to capture the hierarchical document structure and reduce the doc token length for inference efficiency.
\section{Challenges \& Future Directions} \label{sec:challenge}

As discrete tokenizers continue to evolve and find new applications in multimodal understanding and generation, several key challenges and opportunities emerge.

\subsection{Challenges}

\paragraph{Trade-offs Between Compression and Fidelity.}
Existing visual tokenizers (such as VQ-GAN~\cite{esser2021vqgan}) achieve efficient generation by heavily compressing the input resolution (e.g., from 256$\times$256 to 16$\times$16), leading to inevitably loss of fine-grained information. 
On the other hand, increasing the token count can enhance quality, but it results in an exponential increase in sequence length and computational cost.
Striking an optimal balance between compression rate and generation quality remains an open challenge in the field.

\paragraph{Trade-offs Between Understanding and Generation.}
A single visual encoder faces challenges in simultaneously optimizing for both understanding and generation tasks. 
The DeepSeek Janus~\cite{chen2025janus} experiment highlights this issue, showing that when a VQ Tokenizer is used, multimodal understanding performance is significantly lower compared to using a specialized semantic tokenizer. 
To address this, Janus-Pro decouples the visual encoder into specialized tokenizers for understanding and generation. While this approach mitigates the issue, it also increases the model's complexity.

\paragraph{Codebook Collapse and Utilization.} Due to the reliance of discrete tokenizers on a finite-sized codebook, a fundamental challenge in training semantic tokenizers is the codebook collapse problem, where only a small subset of the codebook entries are effectively utilized while others remain "dead" or unused. 
This issue particularly affects VQ-VAE-based approaches and their variants, limiting the model's capacity to capture diverse semantic representations.
Recent lookup-free quantization methods~\cite{mentzer2023fsq}~\cite{yu2023magvit-v2} are emerging to adress this problem.

\paragraph{Cross-Modal Alignment and Consistency.} 
Ensuring consistent semantic alignment between tokens from different modalities presents a significant challenge in multimodal systems. 
Current methods often struggle to maintain semantic consistency when processing multiple modalities simultaneously (e.g. text, image, video), particularly in real-time scenarios. 
In tasks such as video summarization~\cite{li2023blip}, where both temporal and spatial elements need to be aligned, ensuring synchronization across tokens becomes even more challenging.

\paragraph{Integration with Foundation Models.} 
While semantic tokenizers show promise in bridging different modalities with large language models (LLMs), their integration presents several challenges. 
These include aligning token vocabularies with LLM architectures, managing computational overhead, and ensuring effective knowledge transfer between foundation models and tokenized representations.
Additionally, the mixed input of multimodal tokens can introduce distributional discrepancies between modalities,  which can cause instability during training.

\subsection{Future Directions}

\paragraph{Adaptive and Dynamic Tokenization.} Future research should focus on developing more flexible tokenization frameworks that can dynamically adapt to different types of input data and task requirements. This includes investigating methods for automatic vocabulary size adjustment and hierarchical tokenization structures that can capture both fine-grained details and high-level semantic concepts.

\paragraph{Efficient Training and Inference.} Development of more efficient training and inference methods represents a crucial direction for future research. This includes exploring lightweight architectures specifically designed for resource-constrained environments and investigating parameter-efficient adaptation methods while maintaining high-quality representations.

\paragraph{Architectural Innovations.}
Unified tokenization frameworks~\cite{yang2023teal} aim to simplify modality-specific encoders. Byte-level models like BLT~\cite{pagnoni2024blt} bypass tokenization entirely by processing raw bytes, potentially resolving alignment issues.
More innovation can be made in model architecture and problem definition to enable the tokenizer to have better efficiency and generalization.

\section{Conclusion} \label{sec:conclusion}

This survey provides a comprehensive overview of discrete tokenizers, which leverage vector quantization (VQ) techniques to transform continuous multimodal features into discrete semantic representations. 
By evolving from vanilla VQ methods to modern approaches, discrete tokenizers have become fundamental components for bridging different modalities with foundation models. 
Their applications span from enhancing recommendation systems' efficiency to enabling high-quality multimodal generation.
By summarizing the challenges faced by discrete tokenizers, we aim to contribute to the understanding and advancement of this field, 
identifying promising directions in the era of foundation models and multimodal AI.

\bibliographystyle{named}
\bibliography{ijcai24}

\end{document}